\documentclass{extarticle}
\usepackage{PRIMEarxiv}

\usepackage[linesnumbered,vlined,ruled,commentsnumbered]{algorithm2e}

\DontPrintSemicolon

\usepackage{graphicx}
\usepackage{float}
\usepackage{amsmath}
\usepackage{amscd}
\usepackage{enumerate}
\usepackage{enumitem}
\usepackage{amsfonts}
\usepackage{amssymb}
\usepackage[utf8]{inputenc}
\usepackage[T1]{fontenc}
\usepackage{amsthm}
\usepackage{imakeidx}
\usepackage{tikz}
\usepackage{tikz-cd}
\usepackage{color}
\usepackage{xcolor}
\usepackage{dsfont}
\usepackage{scalerel,stackengine}
\usepackage{centernot}
\usepackage{chngcntr}
\usepackage{apptools}
\usepackage{lscape}
\usepackage{nameref}
\usepackage{afterpage}
\usepackage{caption}
\usepackage{subcaption}
\usepackage{hyperref}
\usepackage[all]{xy}
\usepackage{wrapfig}
\usepackage{fancyvrb}
\usepackage{listings}
\usepackage{comment}
\usepackage{mathtools}
\usepackage{booktabs}

\newtheorem{theorem}{Theorem}

\newtheorem{Assumption}{Assumption}

\newtheorem{prop}{Proposition} 
\newtheorem{example}{Example}[section] 
 
\newtheorem{rem}{Remark}

\usepackage{natbib}
\setcitestyle{authoryear}

\newcommand{\E}{\mathbb{E}}

\newcommand{\Var}{\operatorname{Var}}

\title{Residual Balancing for Non-Linear Outcome Models in High Dimensions}
\author{Isaac Meza Lopez \\
Department of Economics, Harvard University\\
\small{isaacmezalopez@g.harvard.edu}}

\normalsize
\begin{document}
\maketitle

\begin{abstract}
We extend the approximate residual balancing (ARB) framework to nonlinear models, answering an open problem posed by \citep{athey2018approximate}. Our approach addresses the challenge of estimating average treatment effects in high-dimensional settings where the outcome follows a generalized linear model. We derive a new bias decomposition for nonlinear models that reveals the need for a second-order correction to account for the curvature of the link function. Based on this insight, we construct balancing weights through an optimization problem that controls for both first and second-order sources of bias. We provide theoretical guarantees for our estimator, establishing its $\sqrt{n}$-consistency and asymptotic normality under standard high-dimensional assumptions.
\end{abstract}

\normalsize

\section{Introduction}

In observational studies, practitioners often rely on the unconfoundedness assumption, which posits that treatment assignments are as good as random given observed covariates. This assumption is at the core of many popular methods for causal inference, including regression adjustment, matching, propensity score weighting, and doubly robust methods \citep[e.g.,][]{rosenbaum1983central, imbens2015causal, hirano2003efficient}. These methods, when applied in low-dimensional settings, allow for the estimation of Average Treatment Effects (ATEs) with attractive statistical properties, including asymptotic normality and efficiency.

However, modern datasets often feature a high-dimensional structure, where the number of covariates \( p \) can exceed the sample size \( n \). High-dimensional covariates arise both from richer data collection processes and deliberate expansions of low-dimensional features (e.g., interactions, polynomial expansions, fixed effects). While high-dimensional data can mitigate issues of unmeasured confounding by capturing more covariate variation, they also pose significant challenges for causal inference. In particular, classical approaches that rely on \( p \)-fixed asymptotics may no longer yield valid inferences, and new techniques are required to handle the biases introduced by high-dimensional regularization methods.  This has motivated a new line of research focused on developing robust methods for causal inference in high-dimensional environments.

Recent developments in high-dimensional causal inference have adapted classical methods to this regime. One prominent approach is the Approximate Residual Balancing (ARB) framework introduced by \citep{athey2018approximate}, which combines regularized regression adjustments with balancing weights. ARB demonstrates that, in high-dimensional linear models, it is possible to achieve \(\sqrt{n}\)-consistent and asymptotically normal inference for the ATE by constructing approximately balancing weights and correcting for residual confounding. This framework eliminates the need for consistent propensity score estimation and focuses directly on addressing linear biases in the outcome model.


While ARB provides strong theoretical guarantees and practical utility for linear models, many real-world applications involve nonlinear relationships between covariates and outcomes. Generalized Linear Models (GLMs) are a natural extension, allowing for nonlinear link functions \( \psi(\cdot) \) that capture more complex outcome structures. In such settings, the standard ARB framework is insufficient, as it does not account for the non-linearities introduced by the link function and its derivatives. Recognizing this limitation, \citep{athey2018approximate} noted the extension to GLMs as a crucial direction for future research.

This paper directly addresses that challenge. We develop a principled extension of the ARB methodology to nonlinear outcome models. We show that a naive application of ARB fails in the nonlinear case because it ignores the curvature of the link function. Our key insight is that controlling for estimation bias requires a second-order correction. This leads to a new optimization problem for constructing balancing weights that simultaneously controls for imbalances in both the covariates and their second moments, weighted by the first and second derivatives of the link function, respectively.

Our main contributions are:
\begin{enumerate}
    \item \textbf{A bias decomposition for nonlinear models:} We derive a fundamental bias-variance decomposition for the treatment effect estimator in the presence of a nonlinear link function. This decomposition explicitly characterizes the bias in terms of first and second-order residual imbalances.

    \item \textbf{A second-order balancing procedure:} Based on our decomposition, we propose a new optimization framework for constructing balancing weights that effectively eliminates asymptotic bias without requiring a correctly specified propensity score model.

    \item \textbf{Complete theoretical guarantees:} We prove that under standard sparsity and regularity conditions, our proposed estimator is $\sqrt{n}$-consistent and asymptotically normal, providing a valid basis for statistical inference on treatment effects in high-dimensional nonlinear models.
\end{enumerate}


\subsection{Related work}

Our work builds on several streams of literature in high-dimensional statistics and causal inference.

\paragraph{High-Dimensional Causal Inference.} The primary challenge in high-dimensional causal inference is obtaining estimators that are not only consistent but also asymptotically normal, enabling the construction of valid confidence intervals. Two main paradigms have emerged to address this.

The first is the \emph{rate double robustness} paradigm, where an estimator relies on both sparsity in the outcome regression and  propensity score model, remaining consistent if at least one is correctly specified. The debiased or double machine learning (DML) framework is a leading example, providing a general recipe for constructing estimators insensitive to first-order errors in these nuisance components \citep{belloni2014inference, farrell2015robust, chernozhukov2017double}. While methods relying on sparsity double robustness like those in \citep{bradic2019sparsitydoublerobustinference, ning2020hdcbps} can accommodate a nonlinear propensity score model (e.g., a GLM for a binary treatment), their framework is fundamentally different from ours. Theirs is an approach that requires estimating both an outcome model and a propensity score model. By contrast, we purposely avoid propensity estimation and achieve orthogonality only with respect to the outcome bias via link-sensitive balancing. In nonlinear links, the Taylor remainder contains a $\psi^{\prime \prime}$ weighted quadratic term that does not vanish under first-order balance alone; consequently, we introduce a second-order guardrail (or impose a beta-min regime with support recovery) to render the remainder $o_p\left(n^{-1 / 2}\right)$. Methodologically, hdCBPS enforces balance through PS estimating equations, whereas our weights directly balance cross-group moments tied to the link's derivatives (e.g., $X^{\top} \psi^{\prime}\left(X \hat{\beta}_c\right)$ ), reflecting our outcome-centric residual-balancing perspective. 

Our work is located in the second paradigm, which achieves consistency by focusing on sparsity in a \emph{single} nuisance component (\emph{model double robustness}) but under stronger sparsity restrictions. We provide valid inference under a sparse outcome model without requiring any specification or estimation of the propensity score model.

\paragraph{Balancing and Calibration Weights.}
A parallel line of research constructs weights to \emph{directly} balance covariates, without relying on a correctly specified propensity model. Entropy balancing calibrates weights to exactly match prespecified moments via a KL projection \citep{hainmueller2012entropy}. Stable balancing weights minimize dispersion subject to exact balance constraints, controlling variance while enforcing mean balance \citep{zubizarreta2015sbw}. Empirical balancing calibration weighting achieves semiparametric efficiency within a calibration family by solving a global moment‑calibration problem \citep{chan2016ebcw}. Kernel balancing extends balance to rich function classes by matching RKHS means \citep{hazlett2020kernel}. Generalized optimal matching (and KOM) unifies matching and weighting as optimization problems that trade off imbalance against variance \citep{kallus2020gom} - finding weights that minimize more general measures of distributional discrepancy. Propensity‑based balancing approaches estimate the score to satisfy balance moments, e.g., CBPS \citep{imai2014cbps} and its extension to continuous treatments, CBGPS \citep{fong2018cbgps}. Finally, \citep{li2018balancing} formalize ``balancing weights'' that target estimands such as ATE/ATT/overlap via propensity‑based reweighting tuned to covariate balance. Relative to these methods, our estimator is a \emph{targeted} balancing scheme where the moment set is \emph{link‑sensitive}—we balance cross‑group moments of $X^\top\psi'(X\hat\beta_c)$ (and a second‑order $\psi''$ term) because these are the objects that control the remainder in high‑dimensional nonlinear plug‑ins. The quadratic penalty on $\|\gamma\|_2$ plays the same role as variance/dispersion control in SBW and GOM/KOM, but our constraints and moments are specialized to the nonlinear transport of control outcomes to the treated group.

\paragraph{Inference for High-Dimensional GLMs.}
Our methodology is closely related to the literature on parameter estimation and inference in high-dimensional Generalized Linear Models (GLMs). Foundational work by \citep{negahban_2012} and \citep{vandeGeer} established consistency and rates of convergence for Lasso-penalized $M$-estimators, which include GLMs, under restricted strong convexity conditions. More recently, a line of work has focused on constructing confidence intervals for individual parameters in high-dimensional GLMs. Methods like the debiased Lasso \citep{JMLR:v15:javanmard14a, javanmard-montanari.debiaslasso} construct approximate inverses of the Hessian to remove regularization bias. While these methods provide valid inference for coefficients, our goal is different: we target an aggregate causal parameter (ATT) and, following ARB, debias the first-stage outcome model by \emph{balancing} the residual structure relevant for the estimand. This perspective connects to augmented minimax linear estimation, which chooses weights to approximate a Riesz representer subject to variance control \citep{hirshberg2021aml}, but here the representer is \emph{cross‑group} and depends on the link’s derivatives.

Our estimator can thus be viewed as a form of debiased nonlinear Lasso tailored to the causal estimand of interest. In \citep{hirshberg2018debiasedinferenceaveragepartial}, the same plug‑in plus weighted‑residual template is used for average partial effects in a single-index model \emph{without} treatment assignment, with weights obtained on a single distribution via an augmented minimax linear program. By contrast, we handle a two-sample transport problem (treated to controls), include first-order \emph{link‑sensitive} cross‑group balance, and introduce an explicit second‑order $\psi''$ guardrail to control the nonlinear remainder—features that are unnecessary in their single‑distribution APE setting but essential here.

\paragraph{Higher-Order Corrections in Semiparametric Inference.}
A crucial feature of our method is the use of a second-order correction in the balancing procedure. This idea has deep roots in the semiparametric efficiency literature. \citep{Robins_2008}, for instance, developed higher-order influence functions to construct estimators for nonlinear functionals that are robust to estimation errors in nuisance parameters. The need for such corrections arises when the parameter of interest is not pathwise differentiable, so that first-order approximations are insufficient to remove asymptotic bias. In the machine learning literature, \citep{mackey2018orthogonalmachinelearningpower} also explored the limitations of first-order orthogonality and the potential need for higher-order adjustments. Our work operationalizes this concept in the context of residual balancing, demonstrating that a second-order Taylor expansion of the bias term leads to an essential correction for achieving valid inference where simpler, first-order balancing would fail, a correction that is essential for inference but requires a stronger sparsity condition on the outcome model. \\

The remainder of the paper is organized as follows. Section 2 lays out our statistical setup and notation. Section 3 develops our main theoretical results for the known-link setting; we derive a key bias decomposition, introduce our second-order balancing procedure, and establish the asymptotic normality of our estimator. In Section 4, we extend this framework to the more flexible single-index model where the link function is unknown. Section 5 evaluates the method's performance through simulation studies and a real-world application. Finally, Section 6 concludes with a summary of our findings.

\section{Setup and Notation}
\label{sec:setup}

Our goal is to estimate average treatment effects (ATEs) in the potential outcomes framework. For each unit in a population, there is a pair of potential outcomes, \((Y_i(0), Y_i(1))\), where \( Y_i(1) \) represents the outcome under treatment, and \( Y_i(0) \) represents the outcome under control. Each unit is assigned to either treatment or control, with the treatment assignment denoted by \( W_i \in \{0, 1\} \). Additionally, each unit is characterized by a vector of covariates \( X_i \in \mathbb{R}^p \), where \( p \) can be large, possibly larger than the sample size \( n \).

For a random sample of size \( n \), we observe the triplet \( (X_i, W_i, Y_i^\text{obs}) \), where:
\[
Y_i^\text{obs} = Y_i(W_i) =
\begin{cases}
    Y_i(1), & \text{if } W_i = 1, \\
    Y_i(0), & \text{if } W_i = 0,
\end{cases}
\]
is the realized outcome corresponding to the treatment assignment. Let \( n_t \) and \( n_c \) denote the number of treated and control units, respectively. The feature matrices corresponding to treated and control units are denoted by \( X_t \) and \( X_c \).

In this work, we focus on estimating the average treatment effect for the treated (ATT):
\[
\tau = \frac{1}{n_t} \sum_{i : W_i = 1} \mathbb{E}[Y_i(1) - Y_i(0) \mid X_i].
\]

Throughout this paper, we adopt the notation: $\|A\|_{\infty }:=\max _{i, j}\left|A_{i j}\right|$ for entrywise $\max ;\|A\|_{\infty \rightarrow \infty}:= \max _i \sum_j\left|A_{i j}\right|$ for the matrix $\ell_{\infty}$ operator norm; and $\|v\|_1,\|v\|_2,\|v\|_{\infty}$ for vector norms. We also make the following assumptions:

\begin{Assumption}[Unconfoundedness]
\label{assu:unconf}
The treatment assignment is independent of the potential outcomes conditional on covariates:
\[
W_i \perp (Y_i(0), Y_i(1)) \mid X_i.
\]
\end{Assumption}

\begin{Assumption}[Generalized Linearity]
\label{assu:linearity}
The conditional outcome models satisfy:
\[
\mu_c(x) = \mathbb{E}[Y_i(0) \mid X_i = x] = \psi(x^\top \beta_c),
\quad
\mu_t(x) = \mathbb{E}[Y_i(1) \mid X_i = x] = \psi(x^\top \beta_t),
\]
for all \( x \in \mathbb{R}^p \), where \( \psi(\cdot) \) is a known link function with \( \psi(\cdot) \in \mathcal{C}_3\).

Furthermore, there exists a deterministic sequence $u_n \rightarrow \infty$ and constants $M_2, M_3>0$ such that, with probability $1-o(1)$,
$$
\sup _{|z| \leq u_n}\left|\psi^{\prime \prime}(z)\right| \leq M_2 \log p, \quad \sup _{|z| \leq u_n}\left|\psi^{(3)}(z)\right| \leq M_3 \sqrt{\frac{n}{\log p}},
$$
and $\max _{i \leq n}\left|X_i^{\top} \beta_c\right| \leq u_n, \max _{i \leq n}\left|X_i^{\top} \hat{\beta}_c\right| \leq u_n$ with probability $1-o(1)$.
\end{Assumption}

We allow for nonlinear links with unbounded second and third derivatives, but the growth must be carefully controlled relative to the statistical complexity of the problem. Specifically, the second derivative's growth is restricted to be at most logarithmic in the number of covariates $p$, a mild condition ensuring the Taylor remainders in our analysis remain asymptotically negligible.

\begin{example}
Any GLM is based on modeling the conditional distribution of the potential outcomes $Y(0)|x$, $Y(1)|x$ in an exponential family. When $\psi$ is a known function, it has a direct connection with the cumulant function $\phi$ of an exponential family. In concrete, whenever
\[P_{\beta^*}(Y(W) \mid x) \propto \exp \left(\frac{Y(W)\left\langle x, \beta^*\right\rangle-\phi\left(\left\langle x, \beta^*\right\rangle\right)}{c(\sigma)}\right)\]
then $\psi(x^\top\beta^*)=\phi^\prime(x^\top\beta^*)$. 
\end{example}

Under these assumptions, the ATT can be expressed as:
\[
\tau = \mu_t - \mu_c, \quad \text{where} \quad
\mu_t = \frac{1}{n_t} \sum_{i : W_i = 1} \psi(X_i^\top \beta_t), \quad
\mu_c = \frac{1}{n_t} \sum_{i : W_i = 1} \psi(X_i^\top \beta_c)
\]
Estimating \( \mu_t \) is straightforward using the sample mean of observed treated outcomes:
\[
\hat{\mu}_t = \bar{Y}_t =\frac{1}{n_t} \sum_{i : W_i = 1} Y_i^\text{obs}.
\]

In contrast, estimating \( \mu_c \) is more challenging, especially in high-dimensional settings where \( p > n \). Addressing this challenge is the primary focus of this paper. Under the Approximate Residual Balancing (ARB) framework, the estimation of \( \mu_c \) is decomposed into two stages. First, a model parameterized by \( \beta_c \) is fitted to capture the strongest signals in the outcome model. This is done using regularized regression methods, such as the lasso, which are well-suited for high-dimensional settings. Second, numerical rebalancing is applied to adjust for residual confounding. Specifically, balancing weights \( \gamma_i \) are computed to align the covariate distributions between treated and control units, correcting for the residual differences in the outcomes not explained by the first-stage model. 

The resulting ARB estimator for \( \mu_c \) can be expressed as:
\[
\hat{\mu}_c = \frac{1}{n_t} \sum_{i : W_i = 1} \psi(X_i^\top \hat{\beta}_c) + \sum_{i : W_i = 0} \gamma_i \left(Y_i^\text{obs} - \psi(X_i^\top \hat{\beta}_c)\right),
\]
The first term \( \psi\left(X_i^\top \hat{\beta}_c\right) \) uses the fitted model to estimate the main effect for the control group, and the second term applies rebalancing weights \( \gamma_i \) to the residuals \( Y_i^\text{obs} - \psi(X_i^\top \hat{\beta}_c) \), ensuring that any leftover signal not captured by the main model is adjusted directly.

This two-stage process ensures that strong predictive signals are captured while residual confounding is addressed numerically, enabling robust estimation of \( \mu_c \) in high-dimensional nonlinear models. This estimator extends the ARB methodology to accommodate nonlinear link functions \( \psi \), for example in GLMs. Another interpretation for this estimator is a form of debiased nonlinear lasso \citep{JMLR:v15:javanmard14a, javanmard-montanari.debiaslasso, hirshberg2018debiasedinferenceaveragepartial}.

Our proposed procedure is then:
\paragraph{Step 0: Sample splitting.} We divide the dataset into two folds: one for estimating the outcome model ( $\hat{\beta}_c$ ) and the other for estimating the balancing weights ( $\gamma$ ). This splitting simplifies the theoretical analysis by ensuring independence between estimation errors.

\paragraph{Step 1: Fit the outcome model using nonlinear lasso.} Estimate \(\beta_c\) in the nonlinear model\footnote{If we have a GLM, we can instead use regularized MLE to better suit the form of exponential families.} by solving:
\[
\hat{\beta}_c = \arg \min_{\beta} \left\{ \sum_{i : W_i = 0} \left(\psi(X_i^\top \beta) - Y_i^\text{obs}\right)^2 + \lambda \|\beta\|_1 \right\},
\]
where \(\psi(\cdot)\) is the link function, \(Y_i^\text{obs}\) represents the observed outcomes, and \(\lambda\) is the regularization parameter.

\paragraph{Step 2: Solve the balancing problem.} Define $W_c(\hat\beta_c) = \operatorname{diag}(\psi^{\prime}(X_c \hat\beta_c))\in\mathbb{R}^{n_c\times n_c}$. Moreover, $W_t^\prime(\hat\beta_c)=\operatorname{diag}\left(\psi^{\prime \prime}\left(X_t \hat{\beta}_c\right)\right)$, and $V_i = \psi^{\prime \prime}\left(X_{c, i}^{\top} \hat{\beta_c}\right) X_{c, i} \otimes X_{c, i}$.

Compute approximately balancing weights \(\gamma\) by solving:
\begin{align*}
  \gamma=&\underset{\tilde{\gamma}}{\arg \min }\quad (1-\zeta)\|\tilde{\gamma}\|_2^2+\zeta\left(\eta\left\|\frac{1}{n_t}X_t^\top\psi^\prime(X_t\hat\beta_c)-X_c^\top W_c(\hat\beta_c)\tilde\gamma\right\|_\infty^2 \right. \\
  &\qquad\qquad\qquad \left.+ (1-\eta)\left\|\frac{1}{n_t}\left(X_t^{\top} \otimes X_t^{\top}\right) \operatorname{vec}(W_t^\prime(\hat\beta_c))-\begin{bmatrix}
        V_1, \cdots , V_n
    \end{bmatrix}\tilde\gamma\right\|_\infty^2\right) \\ 
  & \text {subject to } \sum_{\left\{i: W_i=0\right\}} \tilde{\gamma}_i=1 \text { and } 0 \leqslant \tilde{\gamma}_i \leqslant n_{\mathrm{c}}^{-2/3} \nonumber 
\end{align*}
 where $\psi^\prime(X_t\hat\beta_c)\in\mathbb{R}^n$ is interpreted as component-wise derivative.
\paragraph{Step 3: Estimate the average treatment effect \(\tau\).} Compute the ATE as:
\[
\hat{\tau} = \bar{Y}_t - \left\{ \frac{1}{n_t} \sum_{i:W_i=1} \psi(X_i^\top \hat{\beta}_c) + \sum_{i : W_i = 0} \gamma_i \left(Y_i^\text{obs} - \psi(X_i^\top \hat{\beta}_c)\right)\right\},
\]
where \(\bar{Y}_t\) is the average observed outcome for the treated group.

\paragraph{Step 4: Cross-fitting.} Estimate the outcome model and the balancing weights in the other respective fold and aggregate the resulting ATT. \\

While we focus on the ATT, this framework can be adapted to estimate the Average Treatment Effect on the Untreated (ATU) or the overall Average Treatment Effect (ATE) by defining the target population and constructing analogous balancing weights. For instance, the latter can be estimated as
\[
\hat{\tau}_{ATE} = \left\{ \frac{1}{n} \sum_{i} \psi(X_i^\top \hat{\beta}_t) + \sum_{i : W_i = 1} \gamma_{i,t} \left(Y_i^\text{obs} - \psi(X_i^\top \hat{\beta}_t)\right)\right\} - \left\{ \frac{1}{n} \sum_{i} \psi(X_i^\top \hat{\beta}_c) + \sum_{i : W_i = 0} \gamma_{i,c} \left(Y_i^\text{obs} - \psi(X_i^\top \hat{\beta}_c)\right)\right\}
\]
where $\gamma_{t}$ is given by the solution to
\begin{align*}
  \underset{\tilde{\gamma}}{\min }&\quad (1-\zeta)\|\tilde{\gamma}\|_2^2+\zeta\left(\eta\left\|\frac{1}{n}X^\top\psi^\prime(X\hat\beta_t)-X_t^\top W_t(\hat\beta_t)\tilde\gamma\right\|_\infty^2 \right. \\
  &\qquad\qquad\qquad \left.+ (1-\eta)\left\|\frac{1}{n}\left(X^{\top} \otimes X^{\top}\right) \operatorname{vec}(W^\prime(\hat\beta_t))-\begin{bmatrix}
        V_{1,t}, \cdots , V_{n,t}
    \end{bmatrix}\tilde\gamma\right\|_\infty^2\right) \\ 
  & \text {subject to } \sum_{\left\{i: W_i=1\right\}} \tilde{\gamma}_{i,t}=1 \text { and } 0 \leqslant \tilde{\gamma}_{i,t} \leqslant n_{\mathrm{t}}^{-2/3} \nonumber 
\end{align*}
and we have $W_t(\hat\beta_t) = \operatorname{diag}(\psi^{\prime}(X_t \hat\beta_t))\in\mathbb{R}^{n_t\times n_t}$, $W^\prime(\hat\beta_t)=\operatorname{diag}\left(\psi^{\prime \prime}\left(X \hat{\beta}_t\right)\right)$, and $V_{i,t} = \psi^{\prime \prime}\left(X_{t, i}^{\top} \hat{\beta_t}\right) X_{t, i} \otimes X_{t, i}$.

In the next section, we study the theoretical properties of this procedure designed to obtain asymptotic normality for the ATT.

\newpage

\section{Theoretical guarantees}
\subsection{Bias-Variance Decomposition}
To analyze the performance of the ARB estimator for nonlinear models, we rely on the following bias-variance decomposition for the ARB estimate \( \hat{\mu}_c \):

\begin{prop} \label{prop_decomposition}
Under Assumption \ref{assu:linearity}, then
\begin{align}
\label{decomposition}
|\hat\mu_c-\mu_c|\leqslant\underbrace{\left\|\frac{1}{n_t}X_t^\top \psi^\prime(X_t \beta_c)-X_c^\top W_c(\beta_c)\gamma\right\|_\infty\|\hat\beta_c-\beta_c\|_1+\frac{M_2 \log p}{2}\left(\|X_t\|_\infty^2+\|X_c\|_\infty^2\right) \|\hat\beta_c-\beta\|_1^2}_{Bias} + \underbrace{|\gamma^{\top}\varepsilon_c|}_{Variance} 
\end{align}
 where $\psi^\prime(X_t\beta_c)\in\mathbb{R}^n$, $W_c(\beta_c) = \operatorname{diag}(\psi^{\prime}(X_c \beta_c))\in\mathbb{R}^{n_c\times n_c}$ , $\varepsilon_{i}=Y_{i}(0)-\psi(X_{i}^\top \beta_c)$ is the intrinsic noise, and $\|X_{(\cdot)}\|_\infty$ refers to the entry-wise maximum absolute value.
\end{prop}

This decomposition separates the bias and variance contributions to the error of the estimator \( \hat{\mu}_c \). The first term measures the effect of residual imbalances in covariate distributions, weighted by the error \( \hat{\beta}_c - \beta_c \) in estimating the outcome model. The variance term captures the contribution of the intrinsic noise \( \varepsilon_i \) to the estimator, weighted by the balancing weights \( \gamma \).

The bias-variance trade-off is achieved by carefully selecting the weights \( \gamma \) to balance the two terms in the decomposition. Moreover, we want \( \gamma \) and \( \hat{\beta}_c \) satisfying the following rate conditions:
\begin{align*}
\left\|\frac{1}{n_t}X_t^\top \psi^\prime(X_t \beta_c)-X_c^\top W_c(\beta_c)\gamma\right\|_\infty &= O_p\left(\sqrt{\frac{\log p}{n_c}}\right), \quad
\|\hat{\beta}_c - \beta_c\|_1 = O_p\left(k \sqrt{\frac{\log p}{n_c}}\right),
\end{align*}
where \( k \) is the sparsity level of \( \beta_c \).

The balancing weights \( \gamma \) are chosen to control the bias due to residual confounding by aligning the treated and control covariate distributions. Simultaneously, the regularized estimator \( \hat{\beta}_c \) for the outcome model must achieve the desired \( \ell_1 \)-rate, ensuring that the bias remains small even in high dimensions.

This decomposition and the associated conditions form the foundation for analyzing the performance of the ARB estimator and deriving theoretical guarantees. By ensuring that both conditions are met, our approach achieves robust and efficient estimation of \( \mu_c \) in high-dimensional nonlinear models.

\subsection{Balancing weights for nonlinear models}
Balancing weights \( \gamma \) are constructed to minimize bias and variance while guaranteeing a desired rate condition:
\begin{align}\label{opt_problem}
  \gamma=\underset{\tilde{\gamma}}{\arg \min }\left\{(1-\zeta)\|\tilde{\gamma}\|_2^2+\zeta\left\|\frac{1}{n_t}X_t^\top \psi^\prime(X_t \beta_c)-X_c^\top W_c(\beta_c)\tilde{\gamma}\right\|_\infty^2 \text { subject to } \sum_{\left\{i: W_i=0\right\}} \tilde{\gamma}_i=1 \text { and } 0 \leqslant \tilde{\gamma}_i \leqslant n_{\mathrm{c}}^{-2/3}\right\}  
\end{align}

This ensures that the treated covariate distribution is well-aligned with the control covariate distribution in terms of the nonlinear weights \( \psi'(\cdot) \). Observe that the problem (\ref{opt_problem}) relies on an oracle $\beta_c$, which is unknown. A simple plug-in is not enough, since we would have to account for the estimation error. To make the optimization problem in (\ref{opt_problem}) implementable, we further linearize \( \psi'(X_i^\top \beta) \) around \( X_i^\top\hat{\beta_c} \), introducing second-order derivative terms\footnote{This further expansion to higher-order terms is in line with the strand of literature that uses higher-order corrections to get residuals that decay with faster rates. See \citep{Robins_2008}, \citep{mackey2018orthogonalmachinelearningpower}.} \( \psi''(\cdot) \) that will impose additional constraints on the balancing weights. Theorem \ref{prop_feasible_decomposition} makes explicit this decomposition.

\begin{theorem} \label{prop_feasible_decomposition}
We have the following bound for the deviation of the ATT estimation:
\begin{align}\label{decomposition_feas}
|\hat\mu_c-\mu_c|&\leqslant\left\|\frac{1}{n_t}X_t^\top\psi^\prime(X_t\hat\beta_c)-X_c^\top W_c(\hat\beta_c)\gamma\right\|_\infty \|\hat\beta_c-\beta_c\|_1 + \left\|\frac{1}{n_t}\left(X_t^{\top} \otimes X_t^{\top}\right) \operatorname{vec}(W_t^\prime(\hat\beta_c))-\begin{bmatrix}
        V_1, \cdots , V_n
    \end{bmatrix}\gamma\right\|_\infty\|\hat\beta_c-\beta_c\|_1^2 \nonumber \\
&\qquad\qquad + |\gamma^{\top}\varepsilon_c|    \nonumber  \\
&\qquad\qquad +O\left(\log p\left(\|X_t\|_\infty^2+\|X_c\|_\infty^2\right)\|\hat\beta_c-\beta_c\|_1^2\right)  + O_p\left(\sqrt{\frac{n}{\log p}}\left(\|X_t\|_\infty^3+\|X_c\|_\infty^3\right) \|\hat\beta_c-\beta_c\|_1^3\right)
\end{align}
where $W_t^\prime(\hat\beta_c)=\operatorname{diag}\left(\psi^{\prime \prime}\left(X_t \hat{\beta}_c\right)\right)$, and $V_i = \psi^{\prime \prime}\left(X_{c, i}^{\top} \hat{\beta}\right) X_{c, i} \otimes X_{c, i}$.
\end{theorem}

Thus, we need to balance a `variance' term as well (to correct the bias from using an estimation for $\beta$ in the balancing problem) but not as aggressively as the first-order term. The balancing problem now becomes:
\begin{align*}
  \gamma= \underset{\tilde{\gamma}}{\arg \min }&\quad (1-\zeta)\|\tilde{\gamma}\|_2^2+\zeta\left(\eta\left\|\frac{1}{n_t}X_t^\top\psi^\prime(X_t\hat\beta_c)-X_c^\top W_c(\hat\beta_c)\tilde\gamma\right\|_\infty^2 \right. \\
  &\qquad\qquad\qquad \left.+ (1-\eta)\left\|\frac{1}{n_t}\left(X_t^{\top} \otimes X_t^{\top}\right) \operatorname{vec}(W_t^\prime(\hat\beta_c))-\begin{bmatrix}
        V_1, \cdots , V_n
    \end{bmatrix}\tilde\gamma\right\|_\infty^2\right) \\ 
  & \text {subject to } \sum_{\left\{i: W_i=0\right\}} \tilde{\gamma}_i=1 \text { and } 0 \leqslant \tilde{\gamma}_i \leqslant n_{\mathrm{c}}^{-2/3} \nonumber 
\end{align*}
where the parameter $\eta$ reflects how aggressively we should balance the second-order correction term. To solve this, we propose reformulating the optimization problem for the balancing weights \( \gamma \) from its penalized form into a direct constrained optimization problem:
\begin{align}
  \gamma= \underset{\tilde{\gamma}}{\arg \min } & \quad (1-\zeta)\|\tilde{\gamma}\|_2^2 \label{opt_problem_feasible} \\ 
  \qquad\qquad\quad \text { subject to } \qquad &\left\|\frac{1}{n_t}X_t^\top\psi^\prime(X_t\hat\beta_c)-X_c^\top W_c(\hat\beta_c)\tilde\gamma\right\|_\infty\leqslant K_1\sqrt{\frac{\log(p)}{n_c}} \tag{\(\mathsf{C1}\)} \\
   \qquad &\left\|\frac{1}{n_t}\left(X_t^{\top} \otimes X_t^{\top}\right) \operatorname{vec}(W_t^\prime(\hat\beta_c))-\begin{bmatrix}
        V_1, \cdots , V_n
    \end{bmatrix}\tilde\gamma\right\|_\infty\leqslant K_2 \tag{\(\mathsf{C2}\)}  \\
  \qquad &  |\tilde{\gamma}_i| \leqslant n_{\mathrm{c}}^{-2/3} \tag{\(\mathsf{C3}\)}
\end{align}

and note that we dropped the normalization constraints. We impose the hard constraint $\left\|\frac{1}{n_t}X_t^\top\psi^\prime(X_t\hat\beta_c)-X_c^\top W_c(\hat\beta_c)\tilde\gamma\right\|_\infty\leqslant K_1\sqrt{\frac{\log(p)}{n_c}}$ to achieve the right rate for $\hat\mu_c$ in light of the decomposition in (\ref{decomposition_feas}). Moreover, since the term in the LHS of equation (\ref{constraint_second_order}) is multiplied by $\|\hat\beta_c-\beta_c\|_1^2$ in the fundamental decomposition, it only suffices to be of constant order\footnote{This also justifies why we don't have higher-order correction terms in the balancing problem.}. In other words, a solution to this problem guarantees that
\begin{align*}
\left\|\frac{1}{n_t}X_t^\top \psi^\prime(X_t \hat\beta_c)-X_c^\top W_c(\hat\beta_c)\gamma\right\|_\infty& = O_p\left(\sqrt{\frac{\log p}{n_c}}\right), \\
\left\|\frac{1}{n_t}\left(X_t^{\top} \otimes X_t^{\top}\right) \operatorname{vec}(W_t^\prime(\hat\beta_c))-\begin{bmatrix}
        V_1, \cdots , V_n
    \end{bmatrix}\tilde\gamma\right\|_\infty &= O_p(1), \\
    \quad\quad\|\gamma\|_\infty &= O_p(n_c^{-2/3})\quad 
\end{align*}

\begin{rem}
    As an alternative to imposing a higher-order correction, we can instead make a beta-min assumption on the signal strength. This allows us to debias over only the relevant set of covariates, rather than correcting the full sample Hessian.
The logic is that if the first-stage GLM lasso recovers the true active set $S=\operatorname{supp}\left(\beta_c\right)$, a property guaranteed by a standard beta-min condition \citep[Eq. 2.18]{buhlmann2011statistics} - then it is sufficient to balance only the first-order moments corresponding to the covariates in $S$ \citep[Eq. 2.19]{buhlmann2011statistics}.

This presents a fundamental trade-off. Our proposed second-order constraint \(\mathsf{C2}\) is a design-agnostic safeguard against the linearization error from $\psi^{\prime}\left(X \hat{\beta}_c\right)$ that works without assuming exact support recovery. The beta-min alternative, in contrast, avoids \(\mathsf{C2}\) and is less computationally demanding (balancing only on $S$), but it hinges on a strong signal that can recover $S$ and may be brittle if coefficients are near the selection threshold.

Ultimately, the choice depends on the research setting. In applications with many weak but nonzero effects, the second-order constraint is preferable. When a sparse, strong signal is plausible and model selection is stable, the beta-min approach is a computationally lighter and often sufficient alternative.
\end{rem}

\subsection{Feasibility of the Optimization Problem}

Our next result guarantees that the covariate balancing enjoys fast enough rates by proving that the optimization problem above is feasible for a wide class of designs. This is made concrete in the next assumption.

\begin{Assumption}[Subgaussian Design]\label{assu:tid}
   Let $m_t(\beta) = \E[X_i\psi^{\prime}(X_i\beta)\,|\,W_i=1]$, $\Sigma_c = \operatorname{Var}(X_i|W_i=0)$, $\tilde\Sigma_c(\beta) = \operatorname{Var}(X_i\psi^{\prime}(X_i\beta)|W_i=0)$.  Suppose that we have a sequence of random design problems with $X_{\mathrm{c}}=Q \Sigma_{\mathrm{c}}^{1 / 2}$, where $\psi'(X_i\beta)Q$ has iid entries and is $\varsigma$-subgaussian: 
$$\mathbb{E}\left[\exp \left\{t\left(\psi'(X_i\beta)Q_{i j}-\mathbb{E}\left[\psi'(X_i\beta)Q_{i j}\right]\right)\right\}\right] \leqslant \exp \left(\varsigma^{2} t^{2} / 2\right)$$
, and $X_c, X_t$ are sub-Gaussian with (uniform) $\psi_2$-norm at most $K$. Furthermore, let
$\|m_t(\beta)\|_\infty\geqslant \kappa$, and assume that $\left(\Sigma_{\mathrm{c}}\right)_{j j} \leqslant S$ for all $j=1, \ldots, p$. 
\end{Assumption}

We will condition on $X_c$ when applying Hanson-Wright; the random vector $Q_i$ is independent of $X_c$ and has iid sub-Gaussian entries with $\psi_2$-norm $\varsigma$ . The scalars $\psi^{\prime}\left(X_i \beta\right)$ act as fixed weights in the quadratic forms.

This assumption ensures that the control group covariate matrix \( X_c \) follows a structured design, where the rows are independent sub-Gaussian random vectors scaled by the covariance matrix \( \Sigma_c^{1/2} \). The sub-Gaussian property controls the tail behavior of covariates, while bounds on the variances (\( \Sigma_c \)) and population means (\( m_t(\beta) \)) guarantee that covariate distributions remain well-behaved.

    \begin{theorem}
\label{thm_feasible}
        Under Assumption \ref{assu:tid}, together with  
        $$m_t^\top \tilde\Sigma_{\mathrm{c}}^{-1}\Sigma_{\mathrm{c}}\tilde\Sigma_{\mathrm{c}}^{-1} m_t \leqslant V$$
        for some constant $V>0$. Then the constraints,
        \begin{align}
           \left\|\frac{1}{n_t}X_t^\top\psi^\prime(X_t\hat\beta_c)-X_c^\top W_c(\hat\beta_c)\tilde\gamma\right\|_\infty& \leqslant K_1\sqrt{\frac{\log(p)}{n_c}} \label{constraint_first_order}\\
    \left\|\frac{1}{n_t}\left(X_t^{\top} \otimes X_t^{\top}\right) \operatorname{vec}(W_t^\prime(\hat\beta_c))-\begin{bmatrix}
        V_1, \cdots , V_n
    \end{bmatrix}\tilde\gamma\right\|_\infty &\leqslant K_2 \label{constraint_second_order} \\
   |\tilde{\gamma}_i| &\leqslant n_{\mathrm{c}}^{-2/3} 
        \end{align}
          are feasible w.p. approaching 1 for universal constants $K_1, K_2$.
   \end{theorem}


\subsection{Convergence Rate for the Nonlinear lasso}

We assume the following standard assumptions on the rates for the nonlinear lasso.

\begin{Assumption}[Sparsity] \label{sparsity}
 The parameter vector \( \beta_c \) is \( k \)-sparse, $p$ is at most polynomial in $n$, and it holds that:
\[
\frac{k^2 \log(p)}{\sqrt{n}} \to 0 \quad \text{as } n \to \infty.
\]
\end{Assumption}

\begin{Assumption}[Lasso rate] \label{ass:lasso_rate}
 The $\ell_1$ estimation rate for the parameter $\beta_c$ in Step 1 satisfies
\[\|\hat\beta_c-\beta_c\|_1 \;=\; O_p\!\Big(k\sqrt{\tfrac{\log p}{n_c}}\Big).\]
\end{Assumption}
This assumption is mild and is met in a variety of designs; we record two canonical examples.

\begin{example}[GLM/M-estimation \citep{negahban_2012, vandeGeer}]
\label{ex:glm}
Consider the $\ell_1$ penalized estimator 
\[
\hat\beta_c \in \arg\min_{\beta\in\mathbb{R}^p}\;\Big\{L_n(\beta)+\lambda\|\beta\|_1\Big\},
\qquad
L_n(\beta)=\frac{1}{n_c}\sum_{i:W_i=0}\ell\big(Y_i,\psi(X_i^\top\beta)\big),
\]
with $\lambda\asymp \|\nabla L_n(\beta_c)\|_\infty \asymp \sqrt{\log p/n_c}$.
If (i) the design satisfies a \emph{restricted eigenvalue / restricted strong convexity} condition in a neighborhood of $\beta_c$; (ii) the score concentrates so that $\|\nabla L_n(\beta_c)\|_\infty\lesssim \sqrt{\log p/n_c}$; and (iii) $\beta_c$ is $k$-sparse, then 
\[
\|\hat\beta_c-\beta_c\|_1 \;\le\; C_1\,k\,\sqrt{\tfrac{\log p}{n_c}},
\]
\end{example}
\begin{example}[Isotropic/whitened designs \citep{plan2015generalizedlassononlinearobservations}]
\label{ex:pv}
Let $S_c:=\{i:W_i=0\}$ denote the control indices, and assume the following outcome model on controls,
\[
Y_i=\psi(X_i^\top\beta_c)+\varepsilon_i,\qquad \E[\varepsilon_i\mid X_i]=0,\quad i\in S_c,
\]
with $k$-sparse $\beta_c$. Let the \emph{control design} be sub-Gaussian with covariance $\Sigma_c$. Define $Z_i=X_i^\top\beta_c$ and the Plan--Vershynin constant $\mu:=\frac{\E[Z\,\psi(Z)]}{\E[Z^2]}$.

Suppose the design is isotropic, $\Sigma_c=\E[XX^\top]=I_p$ (or we pre-whiten via $\tilde X:=X\hat\Sigma_c^{-1/2}$ with a consistent $\hat\Sigma_c$).
Consider the \emph{linear} lasso on controls
\[
\hat\theta \in \arg\min_{\theta\in\mathbb{R}^p}\Big\{\frac{1}{n_c}\sum_{i:W_i=0}\!\big(Y_i-X_i^\top\theta\big)^2+\lambda\|\theta\|_1\Big\},
\qquad
\lambda \asymp \sqrt{\frac{\log p}{n_c}}.
\]
Plan--Vershynin show that the population least-squares target equals $\theta^\star=\mu\,\beta_c$ in the isotropic case, and that $Y=X^\top\theta^\star+\xi$ with $\E[X\xi]=0$ and sub-Gaussian $\xi$. Under a standard restricted eigenvalue/compatibility condition and $k$-sparsity, the usual lasso oracle inequalities give
\[
\|\hat\theta-\theta^\star\|_1 = O\left( k\sqrt{\frac{\log p}{n_c}}\right)
\]
If $\mu>0$ (e.g., monotone links), normalizing $\hat\theta$ recovers the index direction with the same rate:
\[
\hat\beta_c:=\frac{\hat\theta}{\|\hat\theta\|_2},
\qquad
\|\hat\beta_c-\beta_c\|_1 = O\left( k\sqrt{\frac{\log p}{n_c}}\right)
\]
The same conclusion holds after pre-whitening, incurring only the usual estimation error of $\hat\Sigma_c^{-1/2}$ in constants.
\end{example}

\subsection{Asymptotic Normality}

Finally, we establish the following asymptotic normality result for the ATT. We make the following assumption on the variance structure 
\begin{Assumption}
    \label{ass:variance}
The intrinsic noise $\varepsilon_{i}=Y_{i}(0)-\psi(X_{i} \beta_{\mathrm{c}})$ is subgaussian and heteroskedastic with $\operatorname{Var}(\varepsilon_{i}|X_i)=\sigma_i^2$. 

Moreover, the dispersion and effective sample size (ESS) satisfies
\begin{enumerate}[label=(\roman*)]
    \item $\|\gamma\|_\infty= O(n_c^{-2/3})$
    \item $\|\gamma\|_2^2 = \Omega_p(n_c^{-1})$ - ruling out superefficiency.
\end{enumerate}
and $s_n^2 = \sum_{i:W_i=0}\gamma_i^2\sigma_i^2$, is nondegenerate :
\[s_n^2\asymp \|\gamma\|_2^2\]
\end{Assumption}

\vspace{3mm}

\begin{theorem}\label{can:result}
    Suppose that the control outcomes $Y_{i}(0)$ are drawn from a
 sparse (Assumption \ref{sparsity}), nonlinear model (Assumption \ref{assu:linearity}), and are independent from treatment assignment (Assumption \ref{assu:unconf}).  Furthermore, suppose that the design satisfies Assumption \ref{assu:tid}, and the $\ell_1$ estimation error of $\beta_c$ in the outcome model satisfies Assumption \ref{ass:lasso_rate}. 
 
 Let the estimate $\hat{\mu_c}$ be
\[\hat{\mu}_c = \frac{1}{n_t} \sum_{i : W_i = 1} \psi(X_i^\top \hat{\beta}_c) + \sum_{i : W_i = 0} \gamma_i \left(Y_i^\text{obs} - \psi(X_i^\top \hat{\beta}_c)\right)\]
with balancing weights, $\gamma$, solving (\ref{opt_problem_feasible}).
Finally, if the variance satisfies Assumption \ref{ass:variance},  then $\hat{\mu}_{\mathrm{c}}$ is asymptotically Gaussian,
\begin{align*}
& (\hat{\mu}_{\mathrm{c}}-\mu_{\mathrm{c}}) /s_n \overset{d}{\longrightarrow} \mathcal{N}\left(0, 1\right) 
\end{align*}
\end{theorem}

Theorem~\ref{can:result} yields an asymptotically linear representation with influence function
\(\varphi_i=\gamma_i\,\varepsilon_i\) over the control sample:
\[
\hat\mu_c-\mu_c
= \sum_{i:W_i=0}\gamma_i\,\varepsilon_i
\;+\; o_p(s_n),
\qquad
s_n^2 := \Var\!\Big(\sum_{i:W_i=0}\gamma_i\,\varepsilon_i \,\Big|\,\mathcal F_n\Big)
= \sum_{i:W_i=0}\gamma_i^2\,\sigma_i^2,
\]
where \(\sigma_i^2:=\Var(\varepsilon_i\mid X_i)\) (so \(s_n^2=\sigma^2\|\gamma\|_2^2\) under homoskedasticity).

The dispersion constraint \(\|\gamma\|_\infty\le n_c^{-2/3}\) together with \(\|\gamma\|_2^2\asymp n_c^{-1}\)
ensures the Lyapunov/Lindeberg condition via \(\|\gamma\|_\infty/\|\gamma\|_2\to 0\) and rules out
superefficiency by preventing a few units from dominating, which implies \(s_n^2\asymp n_c^{-1}\)
when the \(\sigma_i^2\) are bounded away from \(0\) and \(\infty\).
For inference under heteroskedasticity we use the robust estimator
\(\widehat s_n^{\,2}=\sum_{i:W_i=0}\gamma_i^2\,\hat\varepsilon_i^2\), computed on the evaluation fold,
which satisfies \(\widehat s_n^{\,2}/s_n^{2}\to_p 1\).

We do not target the semiparametric efficiency bound for the full unconfoundedness model because we do not orthogonalize with respect to the propensity score. In exchange, we avoid propensity estimation and retain transparent balance diagnostics.

Note that cross‑fitting is sufficient to guarantee conditional independence between \(\gamma\) and \(\varepsilon\); it is not necessary if one establishes an equicontinuity bound for the map \(\beta\mapsto\gamma(\beta)\).  Consider a further linear approximation $$
\gamma_i(\hat{\beta_c}) \varepsilon_i(0)=\underbrace{\gamma_i\left(\beta_c\right) \varepsilon_i(0)}_{\text {main term }}+\underbrace{\left(\gamma_i(\hat{\beta_c})-\gamma_i\left(\beta_c\right)\right) \varepsilon_i(0)}_{\text {remainder }}
$$
we can establish
\[\sum_{i\,:\,W_i=0}\left[\gamma_i(\hat{\beta_c})-\gamma_i\left(\beta_c\right)\right] \varepsilon_i(0)=\left(\hat{\beta_c}-\beta_c\right)^{\top} \sum_{i\,:\,W_i=0} \nabla \gamma_i(\tilde{\beta_c}) \varepsilon_i(0)=o_p(1)\]
, by imposing an ``equicontinuity condition'' on $\nabla\gamma_i(\cdot)$ to control the remainder term.  The same application of Lyapunov's CLT on $\sum_{i\,:\,W_i=0}\gamma_i(\beta_c)\varepsilon_i(0)$, will yield the result given that $\gamma_i(\beta_c) \perp \varepsilon_i(0)$.

\section{Extension}
\subsection{Single-index model}
\label{sec:single-index}
We can relax Assumption \ref{assu:linearity}, allowing for unknown link functions:

\begin{Assumption}[Smooth single-index model]
\label{assu:single-index}
The conditional outcome models satisfy:
\[
\mu_c(x) = \mathbb{E}[Y_i(0) \mid X_i = x] = \psi(x^\top \beta_c),
\quad
\mu_t(x) = \mathbb{E}[Y_i(1) \mid X_i = x] = \psi(x^\top \beta_t),
\]
for all \( x \in \mathbb{R}^p \), where \( \psi(\cdot) \) is an \emph{unknown} link function with second and third derivatives satisfying : $|\psi^{\prime\prime}(x)|\leqslant M_2 \log p\,$,  $|\psi^{\prime\prime\prime}(x)|\leqslant M_3 \sqrt{\frac{n}{\log p}}$, and moreover $\psi \in \mathcal{H}^s(\alpha,L)$, with $s\ge 3$ and where $\mathcal{H}^s(\alpha,L)=\left\{\psi \in \mathcal{C}^k: \sup _{x \neq y} \frac{\left|\psi^{(k)}(x)-\psi^{(k)}(y)\right|}{|x-y|^\alpha} \leq L\right\}$.
\end{Assumption}

Since $\psi$ is unknown, we estimate it nonparametrically along the scalar index $Z=X^\top\beta_c$. To decouple errors, we use a three-fold scheme.

\paragraph{Step 0: Three-fold splitting.} Partition the control sample into disjoint folds $A,B,C$ :

\begin{enumerate}[label=(\Alph*)]
\item  Estimate $\beta_c$ (index).
\item Estimate $\psi',\psi''$ nonparametrically along the index fitted on $A$.
\item Estimate balancing weights $\gamma$ using the estimated derivatives from $B$ and the index from $A$.
\end{enumerate}

\paragraph{Step 1: Fit the outcome model.} On $A$, fit a high-dimensional single-index estimator for $\beta_c$ (GLM penalized MLE \ref{ex:glm} or PV-linear lasso with whitening \ref{ex:pv}), yielding
\[
\|\hat\beta_c-\beta_c\|_2=O_p\!\Big(\sqrt{\tfrac{k\log p}{n_A}}\Big)
\]

\paragraph{Step 2:  Nonparametric derivatives.} On $B$, compute $\hat Z_i=X_i^\top\hat\beta_c$ (fixed given $A$) and estimate $\widehat\psi',\widehat\psi''$ by a one-dimensional smoother of $Y$ versus $\hat Z$ (e.g. local polynomials \citep[Exercise 1.4]{Tsybakov2009}).

Since we have assumed $\psi\in\mathcal{H}^s$, $s\ge 3$, we have the $L_2$ rates \citep{stone_1980}
\begin{align*}
r_{1,n}:=\|\widehat\psi'-\psi'\|_{L_2}=O_p\big(n_B^{-(s-1)/(2s+1)}\big),
\qquad
r_{2,n}:=\|\widehat\psi''-\psi''\|_{L_2}=O_p\big(n_B^{-(s-2)/(2s+1)}\big),
\end{align*}

\paragraph{Step 3: Find balancing weights (with $\ell_2$ balance).} On $C$, define
\[
\widehat W_c=\operatorname{diag}\!\big(\widehat\psi'(X_c\hat\beta_c)\big),\quad
\widehat W_t'=\operatorname{diag}\!\big(\widehat\psi''(X_t\hat\beta_c)\big),\quad
\widehat V_i=\widehat\psi''(X_{c,i}^\top\hat\beta_c)\,X_{c,i}\otimes X_{c,i}.
\]
Let\footnote{The factor $B_X$ arises from lifting pointwise derivative errors to vector gaps via Cauchy–Schwarz: for instance,
\[
\left\|\frac{1}{n_t}\sum_{i:W_i=1}X_i\{\widehat\psi'(Z_i)-\psi'(Z_i)\}\right\|_2
\;\leqslant\;
\sqrt{\frac{1}{n_t}\sum_{i}\|X_i\|_2^2}\,\|\widehat\psi'-\psi'\|_{L_2}
\;\lesssim\; B_X\,r_{1,n}.
\]
and the last inequality follows with high probability given that the first factor concentrates around $\left\{\mathbb{E}\|X\|_2^2\right\}^{1 / 2}=\sqrt{\operatorname{tr}\left(\Sigma_c\right)}$. In practice we should consider a consistent estimate of the variance.} $B_X:=\sqrt{\operatorname{tr}(\Sigma_c)}$. The balancing weights $\gamma$ are chosen by
\begin{align}
\label{eq:l2-balance}
\gamma=\arg\min_{\tilde\gamma}\quad &(1-\zeta)\|\tilde\gamma\|_2^2\\
\text{s.t.}\quad 
&\Big\|\tfrac{1}{n_t}X_t^\top\widehat\psi'(X_t\hat\beta_c)-X_c^\top \widehat W_c\,\tilde\gamma\Big\|_2
\;\leqslant\; K_1\sqrt{\tfrac{k\log p}{n_C}} + c_1 B_X\, r_{1,n},\tag{\(\mathsf{C1}'\)}\\
&\Big\|\tfrac{1}{n_t}(X_t^\top\!\otimes X_t^\top)\operatorname{vec}(\widehat{W_t}')-[\widehat V_1,\ldots,\widehat V_{n_C}]\,\tilde\gamma\Big\|_2
\;\leqslant\; K_2 + c_2 B_X^2\, r_{2,n},\tag{\(\mathsf{C2}'\)}\\
& |\tilde\gamma_i|\leqslant n_c^{-2/3}.\tag{\(\mathsf{C3}'\)}\nonumber
\end{align}
The additive terms $c_1 B_X r_{1,n}$ and $c_2 B_X^2 r_{2,n}$ absorb the derivative‑estimation error when passing from the oracle constraints ($\psi',\psi''$) to the feasible ones ($\widehat\psi',\widehat\psi''$). 
\paragraph{Step 4: Estimate the average treatment effect \(\tau\).} Compute the ATE as:
\[
\hat{\tau} = \bar{Y}_t - \left\{ \frac{1}{n_t} \sum_{i:W_i=1} \widehat\psi(X_i^\top \hat{\beta}_c) + \sum_{i : W_i = 0} \gamma_i \left(Y_i^\text{obs} -  \widehat\psi(X_i^\top \hat{\beta}_c)\right)\right\},
\]
where \(\bar{Y}_t\) is the average observed outcome for the treated group.

\paragraph{Step 5: Cross-fitting.} Rotate folds to aggregate. \\

\begin{rem}
    Estimating $\psi$ requires an index; hence one must either (i) estimate $\beta_c$ first (as above), or (ii) use a joint single‑index method (profile least squares) that alternates between $\beta$ and $\psi$; the three‑fold scheme with a pilot $\hat\beta_c$ keeps the analysis and implementation simpler.
\end{rem}

For the estimator $\hat{\tau}$ to be asymptotically normal, the main term in the error decomposition must be the stochastic component $\left|\gamma^{\top} \varepsilon_c\right|$, while all other bias and remainder terms must be asymptotically negligible. The standard deviation of the main term is $\|\gamma\|_2 \sigma \asymp n^{-1 / 2}$. Therefore, we require all remainder terms to vanish at a rate faster than $n^{-1 / 2}$, i.e., they must be $o_p\left(n^{-1 / 2}\right)$.

This leads to the following sufficient conditions, which formalize the trade-off between the model's sparsity, its dimensionality, and the smoothness of the unknown link function.

\begin{Assumption}[Smoothness \& sparsity] 
\label{assu:sim_tradeoff}
The sparsity level and smoothness of the link function satisfies as $n\to \infty$,

\begin{enumerate}[label=(\alph*)]
    \item $k\,\log p\;n^{-1/2}\to 0$ , 
    \item $\sqrt{ k\, p\,\log p} \;n^{-\frac{s-1}{2 s+1}} \rightarrow 0$ , and
    \item $k\, p\,\log p \;n^{-\frac{4s-3}{4 s+2}} \rightarrow 0$
\end{enumerate}
    
\end{Assumption}

Condition (a) is a standard strong sparsity condition required for high-dimensional inference. Conditions (b) and (c) are new and represent the statistical price of not knowing the link function. They govern how large the dimension $p$ can be relative to the sample size $n$. Conditions (b) and (c) are the most stringent. It shows that for a fixed level of smoothness, the dimension $p$ cannot grow too quickly relative to $n$ for the theory to hold. This is the direct consequence of needing to control the error from the second-derivative estimation term, which involves a $p^2$-dimensional object $(X\otimes X)$.

\begin{theorem}
\label{thm:can-sim}
  Suppose the assumptions of Theorem \ref{can:result} hold, along with the SIM-specific Assumptions \ref{assu:single-index} and \ref{assu:sim_tradeoff}. Let the estimator $\hat\mu_c$ be constructed using the three-fold splitting scheme with an $\ell_2$-balanced weight vector $\gamma$ from Equation \eqref{eq:l2-balance}. Then, the final cross-fitted estimator is asymptotically normal:
 \begin{align*}
& (\hat{\mu}_{\mathrm{c}}-\mu_{\mathrm{c}}) /s_n \overset{d}{\longrightarrow} \mathcal{N}\left(0, 1\right) 
\end{align*}
\end{theorem}



\section{Discussion and Conclusion}

In this paper, we addressed an open problem posed by \citep{athey2018approximate} by extending the Approximate Residual Balancing (ARB) framework to handle high-dimensional nonlinear outcome models. We demonstrated that a naive application of ARB fails in this setting due to bias introduced by the link function's curvature. Our central contribution is the identification of a second-order balancing principle: to achieve valid inference, one must balance not only the covariates weighted by the link's first derivative but also their outer products weighted by the second derivative.

Based on this insight, we developed a practical new estimator and a novel optimization problem for constructing the required balancing weights. We provided rigorous theoretical guarantees, establishing that our estimator is $\sqrt{n}$-consistent and asymptotically normal under standard high-dimensional assumptions. This result holds without requiring any estimation of the propensity score, preserving the key advantage of the ARB paradigm. Furthermore, we extended our framework to the more flexible semi-parametric single-index model, showing that asymptotic normality can be retained at the cost of a stricter trade-off between model sparsity and the link function's smoothness.

In conclusion, this work provides a rigorous and practical solution to a key limitation of the original ARB framework. Our findings help bridge the gap between high-dimensional causal theory and the types of nonlinear models frequently used in empirical research.

\newpage 

\appendix
\section{Proofs}

\subsection{Proof of Proposition \ref{prop_decomposition}}
\begin{proof}
Recall the form of the estimator \(\hat{\mu}_c\), 
\[
\hat{\mu}_c = \frac{1}{n_t} \sum_{\{W_i=1\}} \psi(X_i^\top \hat{\beta}_c) 
+ \sum_{\{W_i=0\}} \gamma_i \left(Y_i^{\text{obs}} - \psi(X_i^\top \hat{\beta}_c)\right),
\]
and
\[
\mu_c = \frac{1}{n_t} \sum_{\{W_i=1\}} \psi(X_i^\top \beta_c).
\]

Using a Taylor expansion for $\psi\left(X_i^{\top} \hat{\beta}_c\right)$ around $X_i^{\top} \beta_c$, we write:
$$
\psi\left(X_i^{\top} \hat{\beta}_c\right)=\psi\left(X_i^{\top} \beta_c\right)+\psi^{\prime}\left(X_i^{\top} \beta_c\right)\left(X_i^{\top}\left(\hat{\beta}_c-\beta_c\right)\right)+R_i,
$$
where the residual term is:
$$
R_i=\frac{1}{2} \psi^{\prime \prime}\left(X_i^{\top} \tilde{\beta}_c\right)\left(X_i^{\top}\left(\hat{\beta}_c-\beta_c\right)\right)^2,
$$
and $\tilde{\beta}_c$ lies on the line segment between $\hat{\beta}_c$ and $\beta_c$.

Substitute this expansion into the estimator $\hat{\mu}_c$ :
\begin{align*}
   \hat{\mu}_c&=\frac{1}{n_t} \sum_{\left\{W_i=1\right\}}\left[\psi\left(X_i^{\top} \beta_c\right)+\psi^{\prime}\left(X_i^{\top} \beta_c\right)\left(X_i^{\top}\left(\hat{\beta}_c-\beta_c\right)\right)+R_i\right] \\
   &\qquad\qquad +\sum_{\left\{W_i=0\right\}} \gamma_i\left(Y_i^{\mathrm{obs}}-\psi\left(X_i^{\top} \beta_c\right)-\psi^{\prime}\left(X_i^{\top} \beta_c\right)\left(X_i^{\top}\left(\hat{\beta}_c-\beta_c\right)\right)-R_i\right) . 
\end{align*}
Thus, 
$$
\begin{aligned}
\hat{\mu}_c-\mu_c= & \underbrace{\frac{1}{n_t} \sum_{\left\{W_i=1\right\}} \psi^{\prime}\left(X_i^{\top} \beta_c\right)\left(X_i^{\top}\left(\hat{\beta}_c-\beta_c\right)\right)}_{\text {Treated Linear Term }}+\underbrace{\frac{1}{n_t} \sum_{\left\{W_i=1\right\}} R_i}_{\text {Treated Residual Term }}+\underbrace{\sum_{\left\{W_i=0\right\}} \gamma_i\left(Y_i^{\text {obs }}-\psi\left(X_i^{\top} \beta_c\right)\right)}_{\text {Residual Variance Term }} \\
& -\underbrace{\sum_{\left\{W_i=0\right\}} \gamma_i \psi^{\prime}\left(X_i^{\top} \beta_c\right)\left(X_i^{\top}\left(\hat{\beta}_c-\beta_c\right)\right)}_{\text {Control Linear Term }}-\underbrace{\sum_{\left\{W_i=0\right\}} \gamma_i R_i}_{\text {Control Residual Term }} .
\end{aligned}
$$

Let $X_t$ and $X_c$ be the covariate matrix for treated and control units. Define the derivative weight matrix as
$$
W_t(\beta_c)=\operatorname{diag}\left(\psi^{\prime}\left(X_t \beta_c\right)\right), \quad W_c(\beta_c)=\operatorname{diag}\left(\psi^{\prime}\left(X_c\beta_c\right)\right)
$$
and the residual vectors as:
\begin{align*}
\varepsilon_c&=Y_c^{\text {obs }}-\psi\left(X_c \beta_c\right)\\
R_c&=\frac{1}{2} \operatorname{diag}\left(\psi^{\prime \prime}\left(X_c \tilde{\beta}_c\right)\right)\left(\left(X_c\left(\hat{\beta}_c-\beta_c\right)\right) \circ\left(X_c\left(\hat{\beta}_c-\beta_c\right)\right)\right)\\
 R_t&=\frac{1}{2} \operatorname{diag}\left(\psi^{\prime \prime}\left(X_t \tilde{\beta}_c\right)\right)\left(\left(X_t\left(\hat{\beta}_c-\beta_c\right)\right) \circ\left(X_t\left(\hat{\beta}_c-\beta_c\right)\right)\right)   
\end{align*}

Then, in matrix notation we can rewrite the terms as
\[\hat{\mu}_c-\mu_c=\underbrace{\frac{1}{n_t} \mathbf{1}_t^{\top}W_t(\beta) X_t\left(\hat{\beta}_c-\beta\right)}_{\text {Treated Linear Term }}-\underbrace{\gamma^{\top} W_c(\beta) X_c\left(\hat{\beta}_c-\beta\right)}_{\text {Control Linear Term }}+\underbrace{\gamma^{\top} \varepsilon_c}_{\text {Residual Variance Term }}+\underbrace{\frac{1}{n_t} \mathbf{1}_t^{\top} R_t-\gamma^{\top} R_c}_{\text {Taylor Residual Terms }} .\]

Finally, by application of Hölder and together with the upper bound of the second derivative of the link function (Assumption \ref{assu:linearity}):
\begin{align*}
    |\hat{\mu}_c-\mu_c|&\leqslant \left\|\frac{1}{n_t}X_t^{\top}W_t(\beta)\mathbf{1}_t-X_c^{\top}W_c(\beta)\gamma\right\|_\infty\|\hat\beta_c-\beta\|_1 + |\gamma^{\top}\varepsilon_c| + \left|\frac{1}{n_t} \mathbf{1}_t^{\top} R_t\right|+\left|\gamma^{\top} R_c\right|\\
     &\overset{\footnotemark}{\leqslant}\left\|\frac{1}{n_t}X_t^{\top}W_t(\beta)\mathbf{1}_t-X_c^{\top}W_c(\beta)\gamma\right\|_\infty\|\hat\beta_c-\beta\|_1 + |\gamma^{\top}\varepsilon_c| +\left\|\frac{1}{n_t}\mathbf{1}_t\right\|_1\|R_t\|_\infty +\left\|\gamma\right\|_1\|R_c\|_\infty \\
      &\leqslant\left\|\frac{1}{n_t}X_t^{\top}W_t(\beta)\mathbf{1}_t-X_c^{\top}W_c(\beta)\gamma\right\|_\infty\|\hat\beta_c-\beta\|_1 + |\gamma^{\top}\varepsilon_c| +\frac{M_2 \log p}{2}\left\|\left(X_t\left(\hat{\beta}_c-\beta\right)\right) \circ\left(X_t\left(\hat{\beta}_c-\beta\right)\right)\right\|_\infty \\
      &\quad \qquad\qquad\qquad\qquad\qquad\qquad\qquad\qquad \qquad\qquad\qquad + \frac{M_2 \log p}{2}\left\|\left(X_c\left(\hat{\beta}_c-\beta\right)\right) \circ\left(X_c\left(\hat{\beta}_c-\beta\right)\right)\right\|_\infty 
\end{align*}
\footnotetext{For general linear lasso-type rates the optimal Hölder inequality to use is the $\infty-1$ pairing. However, we use the $1-\infty$ pairing since our primitive assumption is on the error $\|\hat\beta_c-\beta_c\|_1$ instead of the prediction error (Cf. Section \ref{sec:single-index}). }
Hence,
\begin{align*}
     |\hat{\mu}_c-\mu_c|&\lesssim \left\|\frac{1}{n_t}X_t^{\top}W_t(\beta)\mathbf{1}_t-X_c^{\top}W_c(\beta)\gamma\right\|_\infty\|\hat\beta_c-\beta\|_1 + |\gamma^{\top}\varepsilon_c| + \frac{M_2 \log p}{2} \left(\|X_t\|_\infty^2+\|X_c\|_\infty^2\right) \|\hat\beta_c-\beta\|_1^2 \\
     &= \left\|\frac{1}{n_t}X_t^{\top}\psi^{\prime}(X_t\beta_c)-X_c^{\top}W_c(\beta)\gamma\right\|_\infty\|\hat\beta_c-\beta\|_1 + |\gamma^{\top}\varepsilon_c| + \frac{M_2 \log p}{2} \left(\|X_t\|_\infty^2+\|X_c\|_\infty^2\right) \|\hat\beta_c-\beta\|_1^2 
\end{align*}
\end{proof}
\subsection{Proof of Theorem \ref{prop_feasible_decomposition}}
\begin{proof}
Using a Taylor expansion of the scalar map $z\mapsto \psi'(z)$ around $X_i^\top\hat\beta_c$, we have
\[
 \psi'(X_i^\top \beta_c)
 = \psi'(X_i^\top\hat\beta_c)
 + \psi''(X_i^\top \hat\beta_c)\, X_i^\top (\beta_c - \hat\beta_c)
 + \frac{1}{2}\,\psi'''(X_i^\top \tilde\beta_c)\,\big(X_i^\top(\beta_c - \hat\beta_c)\big)^2,
\]
where $\tilde\beta_c$ lies on the line segment between $\hat\beta_c$ and $\beta_c$.

Therefore
\begin{align*}
  \frac{1}{n_t}X_t^\top\psi'(X_t\beta_c)
  - X_c^\top W_c(\beta_c)\gamma
  &= \bigg[\frac{1}{n_t}X_t^\top\psi'(X_t\hat\beta_c) - X_c^\top W_c(\hat\beta_c)\gamma\bigg] \\
  &\quad + \bigg[\frac{1}{n_t}X_t^\top\operatorname{diag}\!\big(\psi''(X_t\hat\beta_c)\big)X_t
          - X_c^\top \operatorname{diag}\!\big(\psi''(X_c\hat\beta_c)\circ \gamma\big)X_c\bigg](\beta_c-\hat\beta_c) \\
  &\quad + \frac{1}{n_t}X_t^\top\operatorname{diag}\!\big(\psi'''(X_t\tilde\beta_c)\big)\,\{X_t(\beta_c-\hat\beta_c)\}^{\circ 2} \\
  &\qquad\qquad - X_c^\top \operatorname{diag}\!\big(\psi'''(X_c\tilde\beta_c)\circ \gamma\big)\,\{X_c(\beta_c-\hat\beta_c)\}^{\circ 2}.
\end{align*}

Taking $\ell_\infty$ norms and using Hölder/triangle inequalities,
\begin{align*}
   \big\|\tfrac{1}{n_t}X_t^\top\psi'(X_t\beta_c)-X_c^\top W_c(\beta_c)\gamma\big\|_\infty
   &\le \big\|\tfrac{1}{n_t}X_t^\top\psi'(X_t\hat\beta_c)-X_c^\top W_c(\hat\beta_c)\gamma\big\|_\infty \\
   &\quad + \underbrace{\big\|\tfrac{1}{n_t}X_t^\top\operatorname{diag}\!\big(\psi''(X_t\hat\beta_c)\big)X_t
   - X_c^\top \operatorname{diag}\!\big(\psi''(X_c\hat\beta_c)\circ \gamma\big)X_c\big\|_{\infty}}_{(A)}
      \,\|\beta_c-\hat\beta_c\|_1 \\
   &\quad + \underbrace{\big\|X_c^\top \operatorname{diag}\!\big(\psi'''(X_c\tilde\beta_c)\circ \gamma\big)\big\|_{\infty \rightarrow \infty}
         \,\big\|\{X_c(\beta_c-\hat\beta_c)\}^{\circ 2}\big\|_{\infty}}_{(B)} \\
   &\quad + \underbrace{\big\|\tfrac{1}{n_t}X_t^\top\operatorname{diag}\!\big(\psi'''(X_t\tilde\beta_c)\big)\big\|_{\infty \rightarrow \infty}
         \,\big\|\{X_t(\beta_c-\hat\beta_c)\}^{\circ 2}\big\|_{\infty}}_{(C)} .
\end{align*}
Here $\|\cdot\|_\infty$ on matrices is the entrywise maximum (i.e., $\|\mathrm{vec}(\cdot)\|_\infty$), and
$\|\cdot\|_{\infty\to\infty}$ is the operator norm induced by $\ell_\infty$ (max row sum).

We now bound the three terms.

\paragraph{(A)} Using $\|\cdot\|_\infty=\|\mathrm{vec}(\cdot)\|_\infty$ and
$\mathrm{vec}(X^\top \mathrm{diag}(a) X)=(X^\top\!\otimes X^\top)\,\mathrm{vec}(\mathrm{diag}(a))$,
\begin{align*}
\mathrm{vec}\!\Big(X_t^\top\operatorname{diag}\!\big(\psi''(X_t\hat\beta_c)\big)X_t\Big)
&= \big(X_t^\top\!\otimes X_t^\top\big)\,\mathrm{vec}\!\big(W_t'(\hat\beta_c)\big), \\
\mathrm{vec}\!\Big(X_c^\top\operatorname{diag}\!\big(\psi''(X_c\hat\beta_c)\circ\gamma\big)X_c\Big)
&= \sum_{i=1}^{n_c}\gamma_i\,\psi''(X_{c,i}^\top\hat\beta_c)\,\mathrm{vec}\!\big(X_{c,i}X_{c,i}^\top\big) \\
&= \begin{bmatrix} V_1 & \cdots & V_{n_c}\end{bmatrix}\gamma,
\end{align*}
where $W_t'(\hat\beta_c):=\operatorname{diag}(\psi''(X_t\hat\beta_c))$ and
$V_i:=\psi''(X_{c,i}^\top\hat\beta_c)\,(X_{c,i}\otimes X_{c,i})\in\mathbb{R}^{p^2}$.
Hence
\[
(A)=\Big\|\tfrac{1}{n_t}\big(X_t^\top\!\otimes X_t^\top\big)\,\mathrm{vec}\!\big(W_t'(\hat\beta_c)\big)
      - \begin{bmatrix} V_1 & \cdots & V_{n_c}\end{bmatrix}\gamma\Big\|_\infty.
\]

\paragraph{(B)} Let $a_i:=\psi'''(X_{c,i}^\top\tilde\beta_c)\,\gamma_i$. Then
\[
\big\|X_c^\top \operatorname{diag}(a)\big\|_{\infty\to\infty}
= \max_{1\le j\le p}\sum_{i=1}^{n_c}|X_{c,ij}|\,|a_i|
\le \|X_c\|_\infty\,\|a\|_1
\le M_3\sqrt{\tfrac{n}{\log p}}\,\|X_c\|_\infty\,\|\gamma\|_1,
\]
using Assumption~\ref{assu:linearity}. Under our weighting constraints
($\gamma_i\ge 0$, $\sum_{i:W_i=0}\gamma_i=1$) we have $\|\gamma\|_1=1$. Also
\[
\big\|\{X_c(\beta_c-\hat\beta_c)\}^{\circ 2}\big\|_\infty
\le \|X_c(\beta_c-\hat\beta_c)\|_\infty^2
\le \|X_c\|_\infty^2\,\|\beta_c-\hat\beta_c\|_1^2.
\]
Thus
\[
(B) \;\le\; M_3\sqrt{\tfrac{n}{\log p}}\,\|X_c\|_\infty^3\,\|\beta_c-\hat\beta_c\|_1^2.
\]

\paragraph{(C)} Similarly, with $a_i:=\psi'''(X_{t,i}^\top\tilde\beta_c)$,
\[
\Big\|\tfrac{1}{n_t}X_t^\top \operatorname{diag}(a)\Big\|_{\infty\to\infty}
\le \frac{\|a\|_\infty}{n_t}\max_j\sum_{i=1}^{n_t}|X_{t,ij}|
\le M_3\sqrt{\tfrac{n}{\log p}}\,\|X_t\|_\infty,
\]
and
\[
\big\|\{X_t(\beta_c-\hat\beta_c)\}^{\circ 2}\big\|_\infty
\le \|X_t\|_\infty^2\,\|\beta_c-\hat\beta_c\|_1^2.
\]
Therefore
\[
(C) \;\le\; M_3\sqrt{\tfrac{n}{\log p}}\,\|X_t\|_\infty^3\,\|\beta_c-\hat\beta_c\|_1^2.
\]

Combining the bounds above yields
\begin{align*}
\big\|\tfrac{1}{n_t}X_t^\top\psi'(X_t\beta_c)-X_c^\top W_c(\beta_c)\gamma\big\|_\infty
&\le \big\|\tfrac{1}{n_t}X_t^\top\psi'(X_t\hat\beta_c)-X_c^\top W_c(\hat\beta_c)\gamma\big\|_\infty \\
&\quad + (A)\,\|\beta_c-\hat\beta_c\|_1 \\
&\quad + O\!\Big(\sqrt{\tfrac{n}{\log p}}\big(\|X_t\|_\infty^3+\|X_c\|_\infty^3\big)\,\|\beta_c-\hat\beta_c\|_1^2\Big).
\end{align*}

Finally, substitute this bound into the first term of the inequality in Proposition~\ref{prop_decomposition}
(which multiplies this $\ell_\infty$ gap by $\|\hat\beta_c-\beta_c\|_1$). This yields the claimed
decomposition:
the first line contributes the term with
$\big\|\tfrac{1}{n_t}X_t^\top\psi'(X_t\hat\beta_c)-X_c^\top W_c(\hat\beta_c)\gamma\big\|_\infty\,\|\hat\beta_c-\beta_c\|_1$;
the factor $(A)$ above produces the second-order balance term multiplied by $\|\hat\beta_c-\beta_c\|_1^2$;
and the $O(\cdot)$ remainder contributes the third-order term multiplied by $\|\hat\beta_c-\beta_c\|_1^3$.
\end{proof}

\subsection{Proof of Theorem \ref{thm_feasible}}

\begin{proof}
We find some proxy weights $\gamma^*$ as the solution to the following optimization problem:
\begin{align*}
\gamma^*&=\arg \min _\gamma\|\gamma\|_2^2 +\lambda\left(X_c W_c(\beta) \gamma-\frac{1}{n_t}X_tW_t(\beta)\right)\\
&=\frac{1}{n_t} W_c(\beta)X_c \left(X_c^{\top}  W_c(\beta)X_c\right)^{-1} X_tW_t(\beta) 
\end{align*}
but replace the empirical covariance $X_c^{\top}  W_c(\beta)X_c$ by the population covariance $\tilde\Sigma_c(\beta)$, and the empirical mean $\frac{1}{n_t}X_tW_t(\beta)$ with the population mean $m_t$
We claim that $\gamma^{**} = \frac{1}{n_t} W_c(\beta)X_c \tilde\Sigma_c(\beta)^{-1} m_t(\beta)$ will be a feasible solution to the constraints. First note that,
\begin{align*}
    \left(X_c^\top W_c(\beta)\gamma^{**}\right)_j&= \frac{1}{n_t}e_j^{\top}X_c^\top W_c(\beta)^2X_c \tilde\Sigma_c^{-1}(\beta) m_t(\beta) \\
    &=\frac{1}{n_t} e_j^{\top}\left(\Sigma_c^{1 / 2} Q^{\top} W_c(\beta)^2 Q \Sigma_c^{1 / 2}\right) \tilde{\Sigma}_c^{-1}(\beta) m_t(\beta) \;, \qquad  &(X_c=Q\Sigma_c^{1/2}) \\
    &= \frac{1}{n_t} e_j^{\top} \Sigma_c^{1 / 2}\left[\sum_{i:W_i=0} \psi'(X_i\beta)^2 Q_i Q_i^{\top}\right] \Sigma_c^{1 / 2} \tilde{\Sigma}_c^{-1}(\beta) m_t(\beta) \;, \qquad &(Q^{\top} W_c(\beta)^2 Q=\sum_{i:W_i=0} \psi'(X_i\beta)^2 Q_i Q_i^{\top}) \\
    &=\frac{1}{n_t} \sum_{i:W_i=0} \psi'(X_i\beta)^2 Q_iA_j Q_i^{\top}
\end{align*}
where $A_j=\Sigma_c^{1 / 2} e_j^{\top} \Sigma_c^{1 / 2} \tilde{\Sigma}_c^{-1}m_t(\beta) =\Sigma_c^{1 / 2} \tilde{\Sigma}_c^{-1}m_t(\beta)e_j^{\top} \Sigma_c^{1 / 2} $ is a rank-$1$ matrix with Frobenius norm
\begin{align*}
\left\|A_j\right\|_F^2&=\operatorname{tr}\left(\Sigma_c^{1/2} e_j\left(m_t^{\top} \tilde{\Sigma}_c^{-1} \Sigma_c \tilde{\Sigma}_c^{-1} m_t(\beta)\right) e_j^{\top} \Sigma_c^{1/2}\right) \\
&=\left(m_t^{\top} \tilde{\Sigma}_c^{-1} \Sigma_c \tilde{\Sigma}_c^{-1} m_t(\beta)\right) \cdot \operatorname{tr}\left(\Sigma_c^{1/2} e_j e_j^{\top} \Sigma_c^{1/2}\right)\\
&= m_t^{\top} \tilde{\Sigma}_c^{-1} \Sigma_c \tilde{\Sigma}_c^{-1} m_t(\beta)   \cdot \left(\Sigma_c\right)_{j j} \leqslant V\cdot S
\end{align*}
After applying the Hanson-Wright concentration inequality, given that $\psi'(X_i\beta)Q_i$ has independent, sub-Gaussian entries (Assumption \ref{assu:tid}), we conclude that $\psi'(X_i\beta)^2Q_iA_jQ_i$ is sub-exponential. Specifically, there exists universal constants $C_1, C_2$ such that:
$$\mathbb{E}\left[e^{t\left(\psi'(X_i\beta)^2Q_i^{\top} A_j Q_i-\mathbb{E}\left[\psi'(X_i\beta)^2Q_i^{\top} A_j Q_i\right]\right)}\right] \leqslant \exp \left[C_1 t^2 \varsigma^4 V S\right]$$ 
for all $t \leqslant \frac{C_2}{\varsigma^2 \sqrt{V S}}$. Thus, noting that $\mathbb{E}\left[\mathbf{X}_{\mathrm{c}}^{\top}W_c(\beta) \gamma^{**}\right]=m_t$, we find that for any sequence $t_n>0$ with $t_n^2 / n \rightarrow 0$, the following relation holds for large enough $n$ :
$$
\mathbb{E}\left[\exp \left[\sqrt{n} t_n\left(\mathbf{X}_{\mathrm{c}}^{\top}W_c(\beta) \gamma^{**}-m_t\right)_j\right]\right] \leqslant \exp \left[C_1 t_n^2 \varsigma^4 V S\right]
$$

We can turn the above moment bound into a tail bound by applying Markov's inequality. Plugging in $t_n:=\sqrt{\log (p / 2 \delta)} /\left(\varsigma^2 \sqrt{C_1 V S}\right)$ and also applying a symmetric argument to $\left(-\mathrm{X}_{\mathrm{c}}^{\top} W_c(\beta)\gamma^{**}+m_t\right)_j$, we find that for large enough $n$ and any $\delta>0$,
$$
\mathbb{P}\left[\left|\sqrt{n}\left(X_c^{\top}W_c(\beta) \gamma^{**}-m_t\right)_j\right|>2 \varsigma^2 \sqrt{C_1 V S \log \left(\frac{p}{2 \delta}\right)}\right] \leqslant \frac{\delta}{p} .
$$

by applying a union bound,  with probability tending to 1:
\[\left\|X_c^{\top}W_c(\beta) \gamma^{**}-m_t\right\|_{\infty} \leqslant C \varsigma^2 \sqrt{V S \log (p) / n_{\mathrm{c}}}\]
a standard concentration argument with the fact that $n_{\mathrm{t}} / n_{\mathrm{c}} \rightarrow_p \rho$ establishes that, with probability tending to 1 ,
$$
\left\|m_{\mathrm{t}}-\frac{1}{n_t}X_tW_t(\beta)\right\|_{\infty} \leqslant \nu \sqrt{2.1 \rho} \sqrt{\log (p) / n_{\mathrm{c}}}
$$
which establishes (\ref{constraint_first_order}).

The second order constraint can be proved analogously. By sub-Gaussianity of $Q_{i j}$ then with probability tending to 1 
$$\left\|\gamma^{**}\right\|_{\infty} \leqslant n^{-2/3}$$
\end{proof}

\subsection{Proof of Theorem \ref{can:result}}

\begin{proof}
    From the decomposition result \ref{prop_feasible_decomposition}, we have that
     \begin{align*}
\hat\mu_c-\mu_c&=\sum_{i\,:\,W_i=0}\gamma_i(\hat\beta_c)\varepsilon_i(0)\\
 &\qquad\qquad +O_p\left(\left\|\frac{1}{n_t}X_t^\top\psi^\prime(X_t\hat\beta_c)-X_c^\top W_c(\hat\beta_c)\gamma\right\|_\infty \|\hat\beta_c-\beta_c\|_1\right)  \\
&\qquad\qquad + O_p\left(\left\|\frac{1}{n_t}\left(X_t^{\top} \otimes X_t^{\top}\right) \operatorname{vec}(W_t^\prime(\hat\beta_c))-\begin{bmatrix}
        V_1, \cdots , V_n
    \end{bmatrix}\gamma\right\|_\infty\|\hat\beta_c-\beta_c\|_1^2 \right)\nonumber \\
    &\qquad \qquad +O\left(\log p\left(\|X_t\|_\infty^2+\|X_c\|_\infty^2\right)\|\hat\beta_c-\beta\|_1^2\right)  + O_p\left(\sqrt{\frac{n}{\log p}}\left(\|X_t\|_\infty^3+\|X_c\|_\infty^3\right) \|\hat\beta_c-\beta\|_1^3\right)
\end{align*}

Now, since each entry of $X_c, X_t$ are sub-Gaussian by Assumption \ref{assu:tid} : $\operatorname{Pr}\left(\left|X_{i j}\right| \geq t\right) \leq 2 \exp \left(-c t^2 / K^2\right) \forall t \geq 0$
for a universal constant $c>0$. Then by a union bound, $
\operatorname{Pr}\left(\|X\|_{\infty} \geq t\right) \leq 2 n p \exp \left(-c t^2 / K^2\right)$. Therefore, with high probability $
\|X_c\|_{\infty} \leq \frac{K}{\sqrt{c}} \sqrt{\log \left(\frac{2\, n_c\, p}{\delta}\right)} = O\left(\sqrt{\log (n_c\,p)}\right)
$ and similarly for $X_t$.

Theorem \ref{thm_feasible} in conjunction with Assumption \ref{ass:lasso_rate}, and the previous bound gives us that
 \begin{align*}
\hat\mu_c-\mu_c&=\sum_{i\,:\,W_i=0}\gamma_i(\hat\beta_c)\varepsilon_i(0)\\
 &\qquad\qquad +O_p\left(k \frac{\log p}{n_c}\right)  \\
&\qquad\qquad + O_p\left( k^2 \frac{\log p}{n_c}\right)\nonumber \\
    &\qquad \qquad +O_p\left(k^2 \log (n_c\,p) \frac{(\log p)^2}{n_c}\right)+O_p\left(k^3 \big(\log (n_c\,p)\big)^{3/2} \frac{\log p}{n_c}\right) 
\end{align*}
, and by Assumption \ref{sparsity}, all the $O_p$ terms decay faster than $\frac{1}{\sqrt{n_c}}$. We conclude by applying Lyapunov's CLT, to $\sum_{i\,:\,W_i=0}\gamma_i(\hat\beta_c)\varepsilon_i(0)$, using the subgaussianity of $\varepsilon_i(0)$ and noting that by the sample splitting $\gamma_i(\hat\beta_c) \perp \varepsilon_i(0)$:

Denote by $\mathcal{F}_n$ the $\sigma$-field
that includes the sample where $\gamma(\hat\beta_c)$ was estimated. Let $Z_{n, i}:=\gamma_i(\hat{\beta}_c) \varepsilon_i(0)$ and note that $\E\left[\varepsilon_i(0) \mid X_i\right]=0$. Then, conditional on $\mathcal{F}_n$,
$$
\E\left[Z_{n, i} \mid \mathcal{F}_n\right]=\gamma_i(\hat{\beta}_c) \E\left[\varepsilon_i(0) \mid \mathcal{F}_n\right]=\gamma_i(\hat{\beta}_c) \E\left[\varepsilon_i(0) \mid X_i\right]=0
$$
and the family $\left\{Z_{n, i}\right\}$ is independent conditional on the $\sigma$-field (since the $\varepsilon_i(0)$ are independent across $i$, and $\gamma$ is $\mathcal{F}_n-$ measurable).

The conditional variance satisfies
$$
s_n^2:=\Var\left(\sum_{i \,:\;W_i=0} Z_{n, i} \mid \mathcal{F}_n\right)=\sum_{i \,:\;W_i=0} \gamma_i^2 \Var\left(\varepsilon_i(0) \mid \mathcal{F}_n\right)= \sum_{i \,:\;W_i=0} \gamma_i^2\sigma_i^2
$$
Assumption \ref{ass:variance} guarantees that 
$s_n^2 \asymp n_c^{-1}$. Given that the $\varepsilon_i(0)$ are sub-Gaussian, then $\E\left(\left|\varepsilon_i(0)\right|^{2+\delta} \mid \mathcal{F}_n\right) \leq C_{2+\delta}<\infty$ uniformly in $i$ for a universal constant $C_{2+\delta}$. Then
$$
\sum_{i \,:\;W_i=0}\E\left[\left|Z_{n, i}\right|^{2+\delta} \mid \mathcal{F}_n\right] \leq C_{2+\delta} \sum_{i \,:\;W_i=0}\left|\gamma_i\right|^{2+\delta} \leq C_{2+\delta}\|\gamma\|_{\infty}^\delta \sum_{i \,:\;W_i=0} \gamma_i^2=C_{2+\delta}\|\gamma\|_{\infty}^\delta\|\gamma\|_2^2
$$

Therefore,
$$
\frac{\displaystyle\sum_{i \,:\;W_i=0}\E\left[\left|Z_{n, i}\right|^{2+\delta} \mid \mathcal{F}_n\right]}{s_n^{2+\delta}} \leq \frac{C_{2+\delta}\|\gamma\|_{\infty}^\delta\|\gamma\|_2^2}{s_n^{1+\delta/2}}=O\left(\left(\frac{\|\gamma\|_{\infty}}{\|\gamma\|_2}\right)^\delta\right)
$$

Our dispersion constraint $\|\gamma\|_{\infty} \leq n_c^{-2 / 3}$, together with $\|\gamma\|_2^2 \asymp n_c^{-1}$, yields $
\left(\frac{\|\gamma\|_{\infty}}{\|\gamma\|_2}\right)^\delta \lesssim\left(\frac{n_c^{-2 / 3}}{n_c^{-1 / 2}}\right)^\delta=n_c^{-\delta / 6} \longrightarrow 0$. Hence Lyapunov's condition holds conditionally on $\mathcal{F}_n$ :
$$
\frac{1}{s_n} \displaystyle\sum_{i \,:\;W_i=0} Z_{n, i} \overset{d}{\longrightarrow} \mathcal{N}(0,1) \quad \text { given } \mathcal{F}_n .
$$
and we conclude after an application of iterated expectations
\[\frac{\displaystyle\sum_{i \,:\;W_i=0} \gamma_i(\hat{\beta}_c) \varepsilon_i(0)}{s_n} \overset{d}{\longrightarrow} \mathcal{N}(0,1)\]
\end{proof}

\subsection{Proof of Theorem \ref{thm:can-sim}}
\begin{proof} 
Let $b_n:=\|\hat\beta_c-\beta_c\|_2=O_p\big(\sqrt{\tfrac{k\log p}{n_A}}\big)$ and define
\[
g_{1,n}:=K_1\sqrt{\tfrac{k\log p}{n_C}} + c_1 B_X r_{1,n},
\qquad
g_{2,n}:=K_2 + c_2 B_X^2 r_{2,n}.
\]
First note that $B_X \asymp \sqrt{p}$. Then, by the same decomposition argument as in Theorem~\ref{prop_feasible_decomposition} but using $\ell_2$-norms, and assuming that $n_A,n_B,n_C\asymp n_c$
\begin{align*}
|\hat\mu_c-\mu_c|
&\leqslant |\gamma^\top\varepsilon_c|\;+\;
\underbrace{g_{1,n}\,b_n}_{\text{first-order}}
\;+\;
\underbrace{g_{2,n}\,b_n^2}_{\text{second-order}}
\;+\;
O_p\!\Big(\big\{\tfrac{1}{n_t}+\|\gamma\|_\infty\big\}\|X_c(\hat\beta_c-\beta_c)\|_2^2\Big) \\
& = |\gamma^\top\varepsilon_c| \\
&\qquad \qquad + O\left(\frac{k\log p}{n_c}\,+\,\frac{\sqrt{p\: k\;\log p}}{n^{(4s-1)/(4s+2)}}\right) \\
&\qquad \qquad +  O\left(\frac{k\log p}{n_c}\,+\,\frac{p\: k\;\log p}{n^{(3s-1)/(2s+1)}}\right) \\
&\qquad \qquad +  O\left(\frac{k\log p}{n_c}\right)
\end{align*}
Similar to the proof of Theorem \ref{can:result}, it is enough to establish that the remainder terms are $o_p\left(n_c^{-1 / 2}\right)$ to establish a triangular-array CLT. Assumption \ref{assu:sim_tradeoff} guarantees exactly this.
\end{proof}


\newpage
\bibliographystyle{authordate1}
\bibliography{References}  
\end{document}